\PassOptionsToPackage{usenames,dvipsnames}{xcolor}
\documentclass[sigconf]{acmart}

\usepackage{ifthen}
\usepackage{calc}
\usepackage{soul}

\makeatletter
\newcommand\ifnotfloat[1]{\@ifundefined{@captype}{#1}{}}
\newcommand\ifnotfloatelse[2]{\@ifundefined{@captype}{#1}{#2}}
\makeatother

\newcommand{\revnote}[2]{\ifnotfloat{\marginpar{\centering #1 \fbox{#2}}}}
\renewcommand{\revnote}[2]{}

\DeclareRobustCommand{\annote}[3]{\textcolor{#3}{\ifthenelse{\equal{#1}{}}{[\scshape #2]}{[\textsc{#2:} \emph{#1}]} \normalcolor}
	\revnote{\color{#3} \footnotesize}{#2}}

\usepackage[utf8]{inputenc}
\usepackage{dblfloatfix}
\usepackage{hyperref}
\usepackage{tabularx}
\usepackage{xcolor}
\newcolumntype{P}[1]{>{\centering\arraybackslash}p{#1}}

\begin{document}

\copyrightyear{2017}
\acmYear{2017}
\setcopyright{acmcopyright}
\acmConference{MobiQuitous 2017}{November 7--10, 2017}{Melbourne, VIC, Australia}\acmPrice{15.00}\acmDOI{10.1145/3144457.3144478}
\acmBooktitle{the 14th EAI International Conference on Mobile and Ubiquitous Systems: Computing, Networking and Services}
\acmISBN{978-1-4503-5368-7/17/11}

\title{Selective Jamming of LoRaWAN using Commodity Hardware}

\author{Emekcan Aras\textsuperscript{1},  Nicolas Small\textsuperscript{1},  Gowri Sankar Ramachandran\textsuperscript{1}, Stéphane Delbruel\textsuperscript{1}, Wouter Joosen\textsuperscript{1} and Danny Hughes\textsuperscript{1,2}}
\affiliation{%
  \institution{1. imec-DistriNet, KU Leuven, Celestijnenlaan 200A, Leuven, 3001, Belgium}
  \institution{2. VersaSense, Kapeldreef 60, Heverleee, 3001, Belgium}
}

\email{{ firstname.lastname }@cs.kuleuven.be}

\renewcommand{\shortauthors}{E. Aras et al.}

\begin{abstract}
Long range, low power networks are rapidly gaining acceptance in the Internet of Things (IoT) due to their ability to economically support long-range sensing and control applications while providing multi-year battery life. LoRa is a key example of this new class of network and is being deployed at large scale in several countries worldwide. As these networks move out of the lab and into the real world, they expose a large cyber-physical attack surface. Securing these networks is therefore both critical and urgent. This paper highlights security issues in LoRa and LoRaWAN that arise due to the choice of a robust but slow modulation type in the protocol. We exploit these issues to develop a suite of practical attacks based around selective jamming. These attacks are conducted and evaluated using commodity hardware. The paper concludes by suggesting a range of countermeasures that can be used to mitigate the attacks.
\end{abstract}

%
%

\ccsdesc[500]{Security and privacy~Embedded systems security}

\keywords{IoT, Security, LPWAN, LoRa, LoRaWAN}

\maketitle


\section{Introduction}
The Internet of Things (IoT) is becoming a critical element of the information technology landscape and is predicted to grow rapidly in scale over the coming decade. Contemporary IoT applications range from consumer devices, through smart homes to safety critical industrial systems.
Examples of IoT systems that demand strong security support include: smart electricity grids~\cite{smartgrid}, smart cities~\cite{smartcity}, connected fleets of vehicles~\cite{smartcity} and healthcare management~\cite{iothealth}.

This vast diversity of end-devices and applications raises new networking challenges, that have encouraged the development of a rich diversity of network protocols and infrastructures.

Contemporary IoT networks may be classified according to their physical radio layer, the bit rate they can achieve, the power consumption or the communication range of the products.
Networks that operate over a small area or have to exchange a large amount of data will be more inclined to use protocols such as Wi-Fi, Bluetooth, or ZigBee~\cite{tech_comp}. Applications that demand long range, low power, and can tolerate low bit rates will be inclined to use Sigfox, LoRa or other sub-GHz low-power protocols~\cite{desub}.

This paper focuses on LoRa, a radio modulation technology, upon which the popular LoRaWAN network infrastructure is built.
As of today, the LoRa Alliance\footnote{https://www.lora-alliance.org} allows the deployment of both public and private networks.
The LoRa Alliance includes both public and private operators, allowing it to be present in more than fifty countries, and having ongoing collaboration with more than five hundred operators\footnote{http://www.semtech.com/wireless-rf/internet-of-things/lora-applications/networks}.
This popularity, flexibility and freedom of deployment motivates our interest in this technology.
The LoRa physical layer uses unlicensed radio frequency bands and is designed to achieve a transmission range of a few kilometres with an emphasis on low power consumption and low bit rate. The modulation exploits Chirp Spread Spectrum (CSS) techniques, allowing it to be robust against channel noise~\cite{chirpss}.

LoRa however has inherent drawbacks caused by compromises that have been made in its design.
First, the low bit rate of LoRa messages means that the air-time of these messages is long which makes them more prone to collisions.
Second, LoRa messages are susceptible to interference by other synchronous LoRa messages if the signal strength of these messages is higher.
LoRa's defence against these issues is three pronged: i) the low intended transmission rate of LoRa devices limits the probability of collisions, ii) channel hopping spreads messages across channels, reducing the probability of collisions, and iii) LoRa devices can trade data rate for sensitivity, to punch through noise~\cite{georgiou_low_2017}.

In this paper we argue that while these defence mechanisms attempt to minimise the likelihood and impact of accidental collisions, they leave LoRa networks exposed to malicious interference.
By correctly timing LoRa-based jamming messages, it becomes possible to deliberately cause the collisions that the technology seeks to avoid with a high success rate. Beyond that, this paper shows that the long-air time of LoRa messages affords the time to perform more sophisticated attacks, such as selective jamming (only jam one device while leaving the others unaffected), and in some cases a wormhole attack, where two devices that are networked using faster technologies can record, jam, and replay recorded messages over time to prevent alarms from triggering while preserving a facade of normal operation. This allows an attacker to temporarily disable specific LoRa devices or even to eliminate select messages.

The paper is structured as follows.
Section~\ref{sec:bg} presents an overview of the LoRa environment and its related standards, along with contentious issues.
Section~\ref{sec:attacks} explores these coexistence issues and proposes to exploit them via three different and complementary jamming-based approaches.
We detail the design and implementation of these three attacks in Section~\ref{sec:implement} and evaluate them in Section~\ref{sec:evaluation}.
In Section~\ref{sec:real} the applicability of these techniques to real-world use cases is discussed, and a set of practical scenarios maximising the usefulness of these three techniques is proposed.
Section~\ref{sec:mitigation} proposes a set of mitigation techniques based on various abstraction levels to reduce the impact of these attacks.
We then conclude this paper and discuss plans for future work in Section~\ref{sec:conclusion}.


%
%
%
%
%
%





\begin{table}
  \caption{Data rate, spreading factor and maximum frame size in LoRa packets \cite{lora_alliance_lorawan_2015}}
  \label{tab:max_payload}
  \begin{tabularx}{\columnwidth}{c|c|c}
    DataRate & Configuration & Maximum Frame Size (bytes) \\
    \hline
    0 & LoRa: SF12 / 125 kHz & 59 \\
    1 & LoRa: SF11 / 125 kHz & 59 \\
    2 & LoRa: SF10 / 125 kHz & 59 \\
    3 & LoRa: SF9 / 125 kHz & 123 \\
    4 & LoRa: SF8 / 125 kHz & 230 \\
    5 & LoRa: SF7 / 125 kHz & 230 \\
    6 & LoRa: SF7 / 250 kHz & 230 \\
  \end{tabularx}
\end{table}


\section{Background and Related Work}
\label{sec:bg}

\subsection{LPWAN}
\label{sec:lpwan}
Internet-of-Things (IoT) applications are starting to adopt a new family of long range communication technologies because of their low power consumption and low cost. These technologies are part of the Low-Power Wide-Area Network (LPWAN) class of networks. LPWAN is comprised of a set of wireless standards targeted for IoT applications with long range communication (in the order of kilometres) and low data rate requirements~\cite{7945893}. Since IoT applications are expected to last for a long period on a single battery, LPWAN technologies are designed to consume low power. LoRa and Sigfox are the leaders in the LPWAN market. These technologies typically follow a star topology, which means the IoT end-devices with sensors and actuators connect directly with a gateway. This operational model not only minimises deployment complexity, but also promotes further adoption of IoT in the context of smart cities~\cite{7919089} and the Industrial IoT~\cite{7945893}.

Sigfox supports long range and low-power communication through binary phase-shift keying modulation scheme. However, Sigfox does not provide support for private deployments and instead requires users to connect to a licensed Sigfox provider. In contrast, the market model of LoRa is flexible as it enables any customer to setup their own LoRa ecosystem. Furthermore, companies such as COMCAST, KPN, and Actility are deploying public LoRa networks to meet market demands\footnote{https://www.lora-alliance.org/member-list}.

\begin{figure}
  \centering
  \includegraphics[width=\columnwidth]{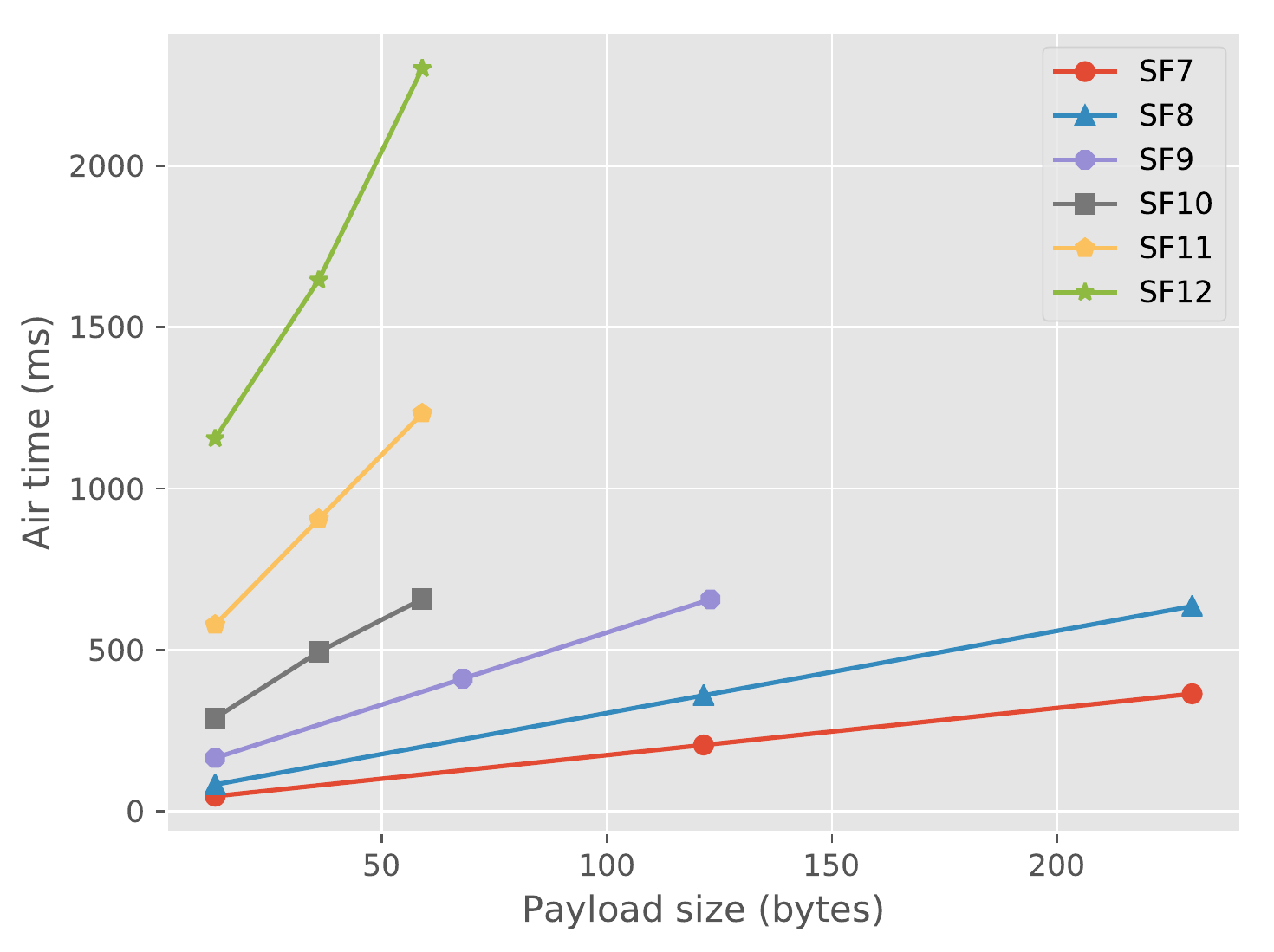}
  \caption{Air time of packets in function of spreading factor (SF) and packet size.}
  \label{fig:airtime}
\end{figure}
\begin{figure*}
  \centering
  \includegraphics[width=\textwidth]{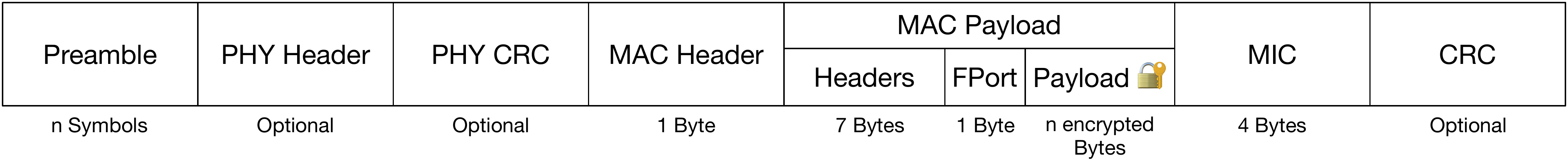}
  \caption{A LoRaWAN Packet.}
  \label{fig:lora_packet}
\end{figure*}
%

\subsection{LoRa}
\label{sec:lora}
LoRa is a long range and low power communication technology based on the Chirp Spread Spectrum (CSS) modulation.
The CSS~\cite{chirpss} modulation scheme uses a chirp signal, which determines spreading across the spectrum. A chirp is a frequency modulated pulse. The frequency can change in a monotonic manner from a lower value to a higher value within a certain time (an upchirp) or  it can change from a higher value to a lower one (a downchirp).
The duration of these chirps is determined by the spreading factor (SF) that determines the sensitivity and transmission speed.
Lower spreading factors enable faster transmissions at the cost of minimum sensitivity, while higher spreading factors result in slow transmissions with greater sensitivity. Figure~\ref{fig:airtime} shows the relationship between spreading factor and time-on-air. In LoRa, \textit{time-on-air} is the time duration of a LoRa signal.
In addition, the data rate of LoRa is determined by the spreading factor. At higher spreading factors, each data bit is represented by multiple chirps, which reduces the data rate. Table~\ref{tab:max_payload}  shows how spreading factor (SF) influences the data rate of LoRa, and the maximum packet size allowed for each SF.

\begin{figure}
  \centering
  \includegraphics[width=0.8\columnwidth]{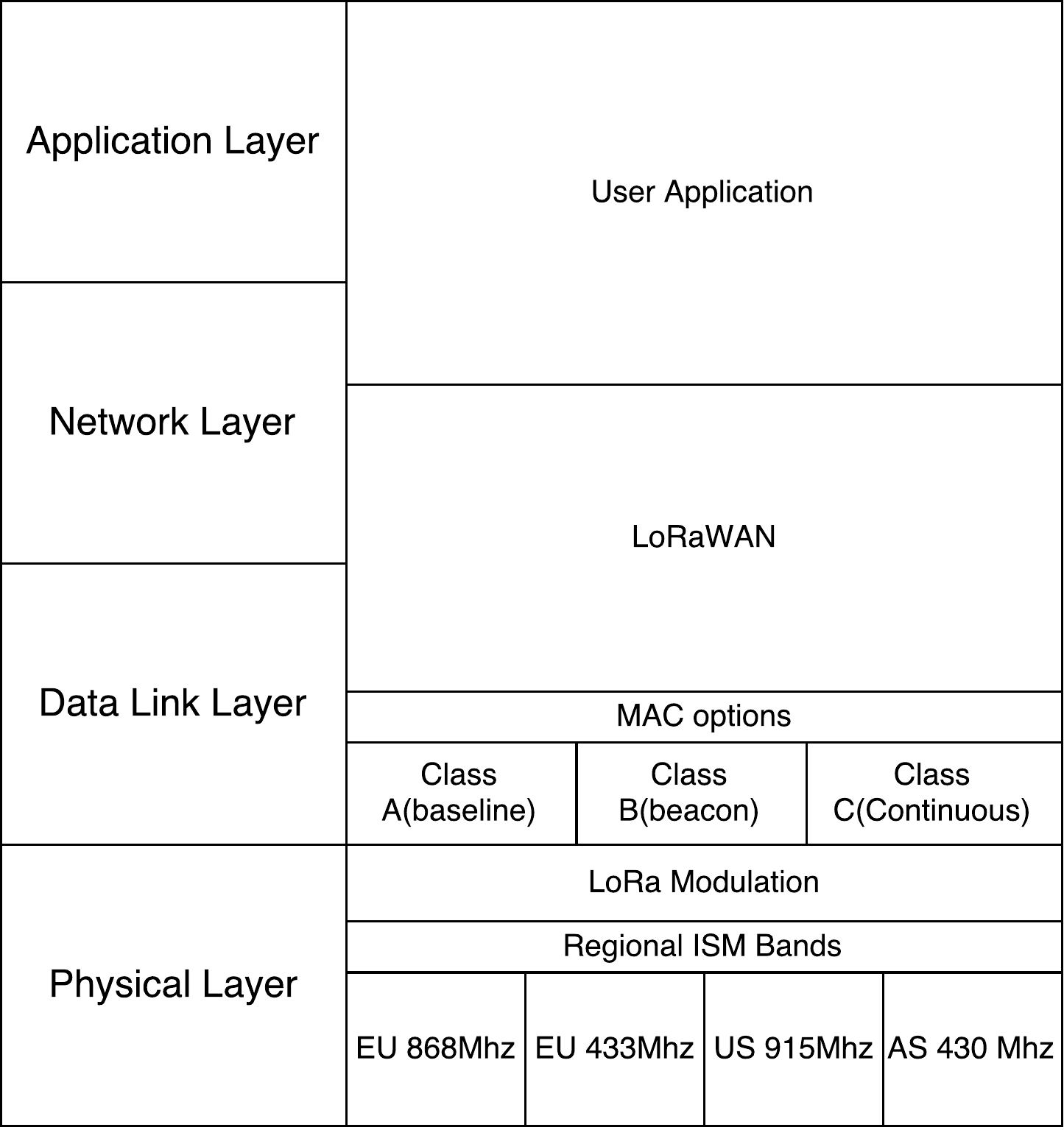}
  \caption{The LoRaWAN stack.}
  \label{fig:lora_stack}
\end{figure}

The LoRa Alliance proposes the LoRaWAN \cite{lora_alliance_lorawan_2015} specification to regulate medium access for LoRa end-devices. LoRa operates at the Industrial, Scientific and Medical (ISM) radio bands, and LoRaWAN defines the operational frequencies in different regions as shown in Figure~\ref{fig:lora_stack}. Furthermore, the telecommunication authority (such as ETSI in Europe and FCC in USA) of the operational region define duty cycling rules for ISM bands. Each LoRa end-device is expected to adhere to the duty cycling regulations.

LoRaWAN follows a star topology, in which all end-devices connect directly with a LoRa gateway.
LoRaWAN classifies the end-devices into three categories: Class A, Class B and Class C (See Figure~\ref{fig:lora_stack}).
Class A end-devices support bi-directional communication, in which each end-device has two short down-link receive windows after an uplink transmission. Class B end-devices also support bi-directional communication,
but they have additional receive windows, which are determined by time-synchronised beacons from the gateway. Finally, Class C
devices allow continuous reception of data due to their maximal receive slots. At the time of writing, only Class A devices are available in the market. In Class A mode, the end-device can transmit messages with or without acknowledgement. Transmissions with acknowledgements are expensive in terms of energy, and may exhaust the duty cycling allowances.

LoRa is a promising technology for IoT, and its architectural model is similar to that of mobile phones, as the public LoRa deployments allow end-devices to connect to a LoRa gateway for a small subscription fee. This operational model coupled with the wide-spread adoption of IoT has increased the number of LoRa applications. For example, \cite{7945359} presents the application of LoRa in a health-care IoT scenario for monitoring blood and vaccine supplies in DR Congo. In addition, LoRa is widely used in applications such as fire detection, radiation leak detection and home security\footnote{http://www.semtech.com/wireless-rf/internet-of-things/lora-applications/briefs}. Given the growing interest in LoRa, and the criticality of these applications, it is important to ensure the security of LoRa communications.


\subsection{LoRaWAN Packets}
\label{sec:lorapacket}
The structure of LoRaWAN packets is defined by the LoRaWAN specification and is depicted in
Figure~\ref{fig:lora_packet}.
As observed, the transmission is initiated by a preamble then followed by radio link layer packets wrapped around MAC packets.
Inside these MAC packets are the data used by the upper layers of this model such as encryption integrity related data or application data.
The hierarchy of that stack is illustrated in Figure~\ref{fig:lora_stack}.
The preamble is used by receiving devices to detect and \emph{lock} on a LoRa signal and is composed of a variable number of upchirps, which depends on the modulation settings, followed by two upchirps and two and a quarter downchirps.
At this stage, the receiver is only aware of a device emitting a LoRa frame and unable to differentiate which device is emitting.
This preamble synchronises the receiver with the reception offset as detailed in~\cite{goursaud_dedicated_2015}, and is a key element to detect precisely when the message part of a LoRa frame will begin to be transmitted.
The preamble is critical, as the LoRa standard operates in the ISM radio bands, shared with license-free communication applications, where detecting the emission of a particular frequency is not enough to identify any transmission.

The MAC layer sits above the physical layer, and is responsible for mediating access to the channel as well as encryption of application data following the specification described in~\cite{lora_alliance_lorawan_2015}.
The MAC layer is first composed of a MAC header stating the message type. It is then followed by the MAC layer payload, which embeds the encrypted application payload.
The physical layer payload is then terminated by the first four bits of the Message Integrity Code (MIC). The message is signed with AES-128 CMAC procedures, following~\cite{chirpss} as indicated in~\cite{lora_alliance_lorawan_2015}.
It is important to note that the messages of the MAC layer are signed by a dedicated key known as the NetworkSessionKey (NwkSKey).  LoRaWAN derives NwkSKey and Application Session Keys (AppSKey) from 128-bit AES key known as the Application Key (Appkey). The AppSKey is used for encrypting and decrypting the payload of application data. LoRaWAN creates a key stream using NwkSKey, AppSKey, and the up-link or down-link frame counter. Therefore, each message is encrypted by using the XOR operation with the corresponding key from the key stream to generate the encrypted payload.
The MAC payload is comprised of the FrameHeader (FHDR), a port field, and then the encrypted application payload.

\emph{The FHDR is of special importance for our work, since its first four bytes represent the end-device address (DevAddr). Furthermore, this part of the MAC payload is not encrypted, which makes LoRaWAN packets susceptible to selective jamming attacks.}




\begin{figure}
  \centering
  \includegraphics[width=\columnwidth]{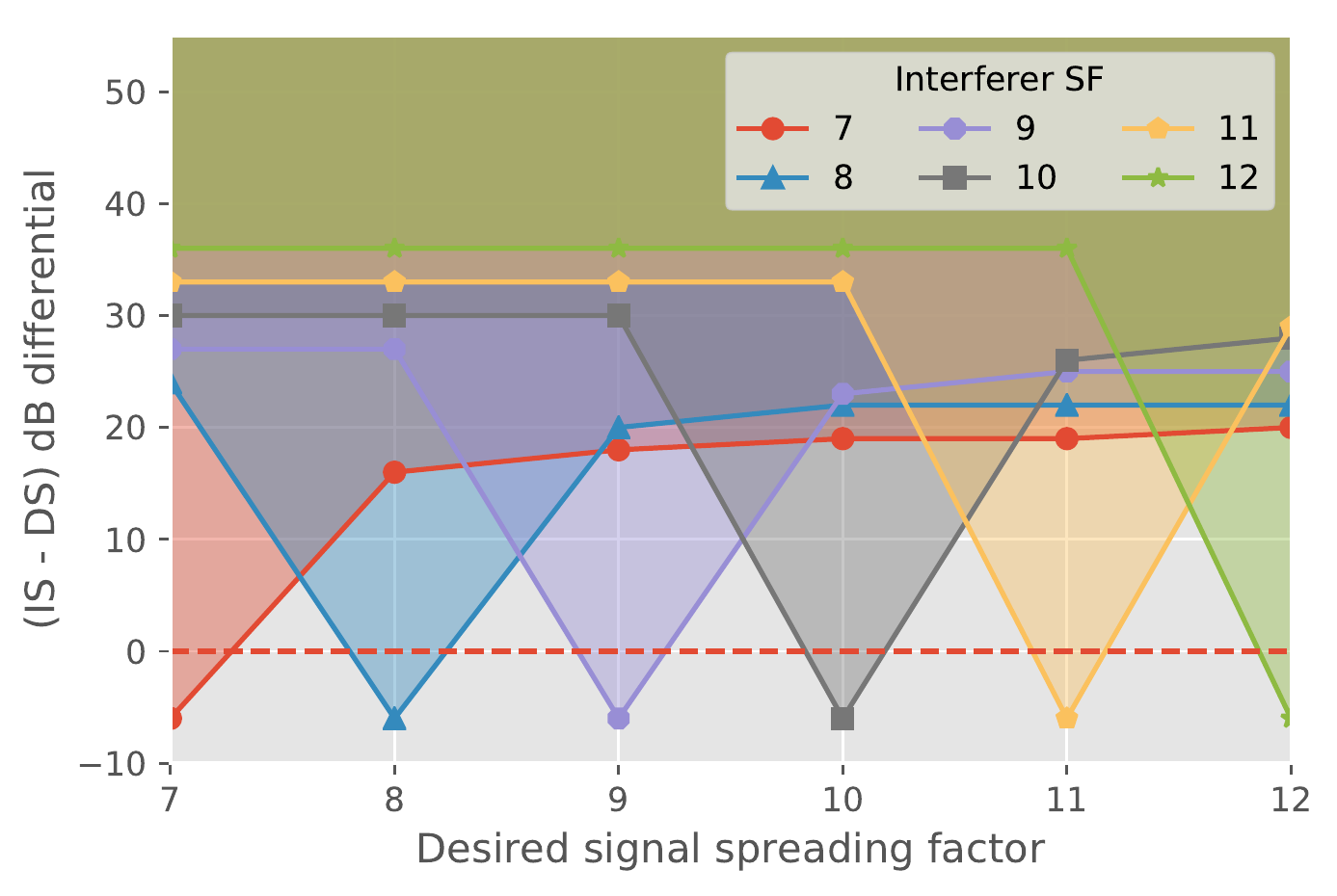}
  \caption{Calculated co-channel rejection thresholds for all combinations of Desired Signal (DS) and Interfering Signal (IS) spreading factors. In each case, if the result of the operation $IS_{strength} - DS_{strength}$ is above the relevant threshold (i.e., inside the coloured areas), the desired signal will be lost. Adapted from \cite{goursaud_dedicated_2015}.}
  \label{fig:db_thresh}
\end{figure}

\begin{table*}[!b]
  \caption{LoRa messages as recorded in the radio module FIFO right after receiving, with and without jammer present.}
  \label{tab:messages}
  \begin{tabularx}{0.8\textwidth}{c|cccc|ccc|c|cccc|cc|cc|c}
    Type & \multicolumn{4}{c|}{Device Address} & \multicolumn{3}{c|}{Frame Headers} & FPort & \multicolumn{4}{c|}{Payload} & \multicolumn{2}{c|}{MIC} & \multicolumn{2}{c|}{CRC} & Jammer Active?  \\
    \hline
    40 &  63 & 56 & 34 & 12  &  00 & 00 & 00 & 01  &  40 & D2 & 83 & 92  &  A8 & C8  &  EB & F3  & No \\
    40 &  63 & 56 & 34 & 12  &  00 & 00 & 00 & 01  &  40 & D2 & 5D & F6  &  71 & DA  &  EB & CB  & Yes \\
    40 &  63 & 56 & 34 & 12  &  00 & 00 & 00 & 01  &  40 & D2 & 3D & 9A  &  1A & 7A  &  C7 & 99  & Yes \\
    40 &  63 & 56 & 34 & 12  &  00 & 00 & 00 & 01  &  40 & D2 & 34 & 90  &  D6 & F5  &  FF & 69  & Yes  \\
  \end{tabularx}
\end{table*}

\subsection{Coexistence Issues in LoRa}
\label{sec:coexistence}

Related work on LoRa focuses on two aspects of LoRa message collisions; the impact of collisions, and their likelihood. As the LoRaWAN protocol does not rely either on channel sensing nor on time synchronisation for collision avoidance \cite{lora_alliance_lorawan_2015}, its main defence against collisions is the low data-rate of the end-devices on the network, making these collisions unlikely\cite{lora_coexistence_new}. This assumption has been found to be far from robust in large scale LoRa deployments however, with coexistence issues being the focus of several papers \cite{reynders_range_2016, georgiou_low_2017}. Packet loss is worse when bi-directional communications are used \cite{pop_does_2017}, although this is also linked to duty cycle limitations being exhausted by ACKs rather than just collisions.

LoRa packet collisions do not always result in packet loss, as the use of CSS enables LoRa gateways to receive simultaneous messages if these are at a different spreading factor (SF), and if these messages are received at a similar power. These collisions matter however when either the messages share the same SFs (the signals are summed and both lost), or one message is transmitted with significantly more power than the other. In the latter case, the more powerful signal is received, while the weaker is disrupted and lost (this also applies to the same SF case) \cite{goursaud_dedicated_2015}. Figure \ref{fig:db_thresh} illustrates this by showing the thresholds above which interfering LoRa signals overpower other LoRa signals, for each SF combination.

\subsection{LoRa Jamming}

Jamming is well studied in many radio technologies, such as Wi-Fi, Bluetooth, Zigbee, etc \cite{mathy, jamming_rfid, jamming_zigbee, jamming_sensornet}.
By abusing the coexistence issues described above, it is possible to jam LoRa messages using well timed malicious transmissions. Previous work on this topic \cite{emekcan_aras_exploring_2017} showed the long air-time of LoRa messages made this triggered jamming (as opposed to continuous) possible and effective. The setup described there jams any LoRa message broadcast on the frequency the jammer is listening to.
This paper proposes to improve this approach by extending it with selective jamming capabilities. Selective jamming requires the classification of messages as they are on-air. By classifying the message, a malicious jammer can systematically jam either a particular type of message or all messages coming from a particular end-device \cite{proano_selective_2010}. As discussed in Section~\ref{sec:lorapacket}, LoRaWAN message headers are not encrypted, which enables selective jamming of LoRaWAN transmissions.




\begin{figure*}
  \centering
  \includegraphics[width=\textwidth]{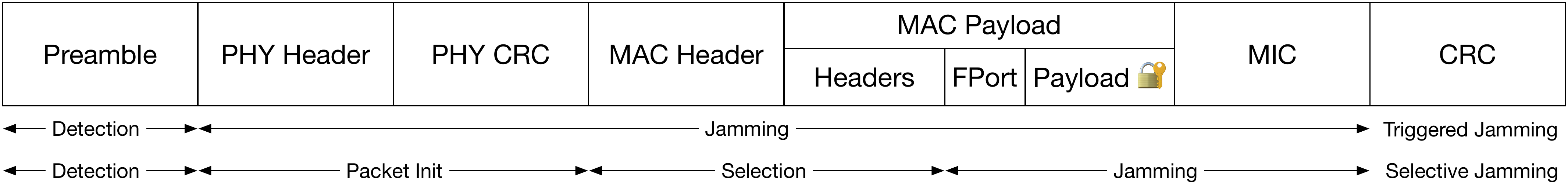}
  \caption{Timing of Triggered and Selective Jamming.}
  \label{fig:timing}
\end{figure*}

\section{LoRa Jamming Attacks}
\label{sec:attacks}

As described Section \ref{sec:bg}, the LoRa physical layer suffers from coexistence issues. LoRa devices which send data simultaneously using certain frequencies and parameters can corrupt each other's signal. By abusing this vulnerability (i.e., increasing the chance of simultaneous transmission from random to guaranteed), it is possible to jam LoRa messages maliciously.

Table \ref{tab:messages} shows four identical LoRaWAN messages as seen by the receiver during the selective jamming experiment detailed in Section \ref{sec:evaluation}. In the last three cases, the jammer is active and corrupts the message, changing the last bits of the message. As can be seen, the first 11 bytes of the packet are not corrupted even when the jammer is active due to the selection time (more of the message should be corrupted in the triggered jamming case). These changed bits cause the message to fail the cyclic redundancy check (CRC), causing the message to be rejected before high-level interpretation by the gateway.
This section describes three different techniques to perform this jamming and highlights their requirements and applications.

\subsection{Triggered Jamming}
In order to avoid simultaneous transmissions, LoRa radio modules have the capability to scan a certain channel to detect whether there is an ongoing LoRa transmission or not (although the use of this capability is not required by the protocol). This capability can also be abused by attackers to detect activity on the channel. Once a LoRa transmission is detected on the channel, the malicious LoRa device can start transmitting in order to jam this transmission.

We tested this triggered jamming technique in our previous work \cite{emekcan_aras_exploring_2017}. \emph{For each SF, we sent 100 messages on the 868.1 MHz frequency and only 3 messages out of 600 (0.5\%) reached the gateway when the triggered jammer was on.}
This vulnerability in the LoRa physical layer allows malicious entities or third parties to use off-the-shelf LoRa devices to increase packet loss in a specific network. In addition, the triggered jammer described here provides a good base for the development of more sophisticated jamming techniques, such as Selective Jamming.

\subsection{Selective Jamming}

Selective jamming is arguably the most sophisticated and efficient jamming technique \cite{mathy}. Since triggered and continuous jamming  \cite{cont_jamming} affect all the devices on a certain frequency, they are easily detectable and simple countermeasures against these  techniques can be taken \cite{sel_vs_cont}. For instance, a particular frequency or network can be marked as jammed, or network administrators can take action, such as changing the communication frequency or enabling channel hopping. Selective jamming on the other hand jams only selected devices or messages, and since the other devices or messages in the network are not jammed, can be much harder for the operator to decide if a device is being jammed or some other technical problem has occurred.

Triggered jamming simply relies on detecting preamble symbols, and then jamming the device without demodulating or decoding any other part of the signal. Selective jamming requires some part of the message to be received and read before action is taken. The decision to jam or not is then made by applying a jamming policy to that content. The policies could consider any of the message headers: message type, device address, frame counter etc. This technique has strict timing requirements that need to be achieved in order to perform accurately. The jamming policy is only concerned with the content following the physical headers. Thus, the time remaining to successfully jam the packet is shorter in selective jamming. The difference in timing requirements between triggered and selective jamming is shown in Figure~\ref{fig:timing} and also detailed in Section~\ref{sec:limits}.

\subsection{Combining Selective Jamming With A Novel Wormhole Attack}
\label{sec:wormhole}

A classic wormhole attack requires two malicious devices amongst a network (E.g., a sensor network). One of the malicious devices receives normal network messages and tunnels them to the other device via a low-latency link. The other device then receives these messages and replays them in a different part of the network to enact an attack~\cite{wormhole_1}. Generally, this type of attack is used to create false route information and routing loops to waste the energy of networks that use mesh topologies \cite{wormhole2}. However, in a LoRaWAN network, the nodes are typically in a star-of-stars topology with gateways forming a transparent bridge. Thus, the classical form of the wormhole attack is not suitable for LoRaWAN networks. As mentioned previously, the LoRaWAN protocol offers a mechanism to prevent replay attacks. Once the MIC of a message is validated by a gateway, any further occurrences of the same sequence number will be rejected.

There is no time related information in LoRaWAN message headers, and only loose timing requirements due to LoRa's long transmission times.
This means that if a message can be recorded and prevented from reaching a gateway, it can be replayed at a later time and appears as a legitimate message, as long as no message with a higher sequence number has been received by the gateway.
The single device selective jamming attack described above cannot achieve this alone, as conventional radio transceivers cannot simultaneously send and receive. This jam and replay attack therefore needs two devices: a sniffer and a jammer.
The sniffer receives messages and decides whether to jam as per normal selective jamming. If the decision to jam is made, it signals to the jammer via a low-latency link, and the jammer immediately jams the message. Unlike the selective jamming attack, the sniffer keeps listening to the original transmission and stores it for later use in a replay attack.

The two devices have to be kept far enough apart so the jammer does not jam the sniffer and therefore prevent the recording of the original message. Ideally then, the sniffer is close to the source device, and the jammer close to the gateway to maximise received signal strength in both cases.
This attack can be used in practice to hide changes in a sensor's state: normal operation messages are jammed and recorded for a while, then during the activity that is to be hidden the sensor messages continue to be jammed. During both phases, the normal operation messages are replayed at a delayed rate to keep apparent sensor values in normal ranges. This simulates packet loss but normal operation otherwise.
The timing requirements of this attack are even more strict than selective jamming due to the communications latency between the two devices.

%


\section{Implementation}
\label{sec:implement}
The attacks described in the previous section were implemented using cheap commodity hardware for their evaluation.
This section presents the details of each of these implementations.

\subsection{Selective Jamming}

In order to selectively jam LoRaWAN messages, a jammer needs to be able to perform the following.

\begin{enumerate}
  \item Detect a LoRaWAN packet.
  \item Start receiving that packet.
  \item Abort receiving if the received content triggers the jamming policy.
  \item Immediately jam the channel.
\end{enumerate}

%

\begin{figure}
  \centering
  \includegraphics[width=\columnwidth]{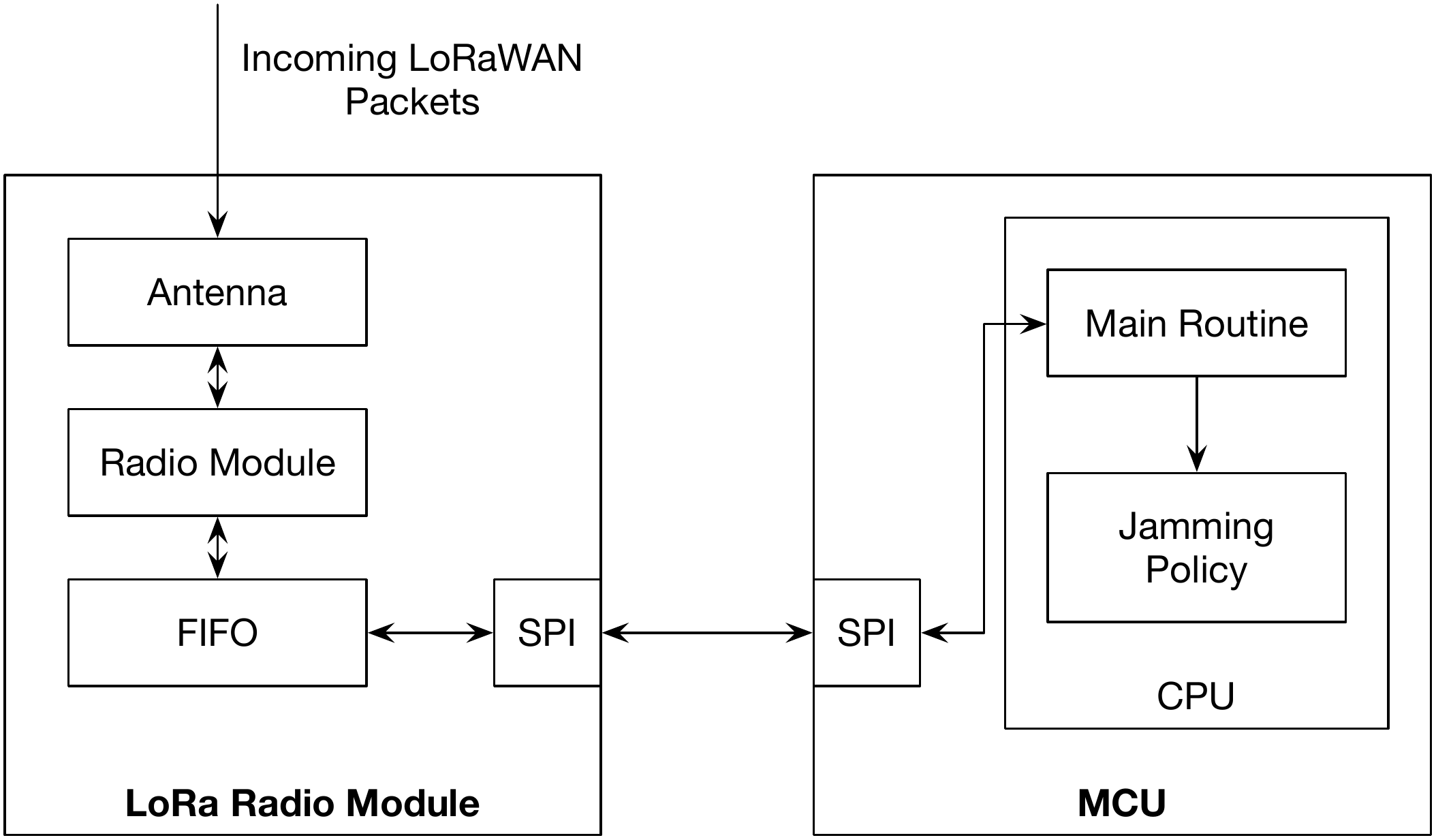}
  \caption{Selective Jammer Architecture.}
  \label{fig:jammer}
\end{figure}

We suggest a simple architecture based on commodity hardware to implement this technique. This architecture is shown in Figure \ref{fig:jammer}.
To implement this architecture, we needed a radio module capable of sending and receiving LoRa packets and software to cycle through these steps.
Some of the LoRa radio modules in the market only support limited, high-level, sets of commands via serial interface. However, selective jamming requires low-level configuration to be done, to allow reading of a message while it is being received and to perform jamming. There are only two devices on the market that allow us to do these low-level configurations: the Semtech Sx1276\footnote{http://www.semtech.com/images/datasheet/sx1276\_ 77\_ 78\_ 79.pdf} and the Hope RFM95/96\footnote{http://www.hoperf.com/upload/rf/RFM95\_ 96\_ 97\_ 98W.pdf} radio modules. Both radio modules communicate via SPI interface. Therefore, any micro-controller board that has an SPI interface can be used to configure these modules. We used the RFM95 radio module together with an Arduino based micro-controller to build our prototype jammer.

There is no support to read LoRaWAN messages byte by byte in the radio module. However, this can be still accomplished by reading the module's FIFO buffer since the radio module uses DMA to write messages byte by byte to memory. The RFM95 radio module has a register that indicates the modem status. It sets a bit in the modem register to indicate that a LoRaWAN physical header has been fully received. Once the flag is set, the LoRaWAN message in the FIFO can be read byte by byte, enabling the selection part of the attack.

During the attack, the radio module starts in receiver mode and waits to receives LoRa modulated signals.
Once a message is detected, if its physical headers are correct, the module starts to write data to the FIFO buffer starting with the message type and device address. The FIFO buffer is then read by software byte by byte. Once enough bytes are read (5 bytes are enough to reach the device address), the jamming policy is applied. If the message is to be jammed, the device switches to jammer mode. Once jamming is done, it switches back to receiver mode again.

\subsection{Combined Selective Jamming and Wormhole Attack}

As described in Section \ref{sec:wormhole}, this attack requires two separate devices, a sniffer and a jammer, which need to perform the following operations, respectively:

\subsubsection*{\bf For the sniffer:}
\begin{enumerate}
  \item Detect a LoRaWAN packet.
  \item Start receiving and recording that packet.
  \item Send a signal to the jammer if received content is enough to trigger the jamming policy.
\end{enumerate}

\subsubsection*{\bf For the jammer:}
\begin{enumerate}
  \item Wait for a signal from the sniffer.
  \item Turn the jammer on once the signal is received.
\end{enumerate}


%


\begin{figure}
  \includegraphics[width=\columnwidth]{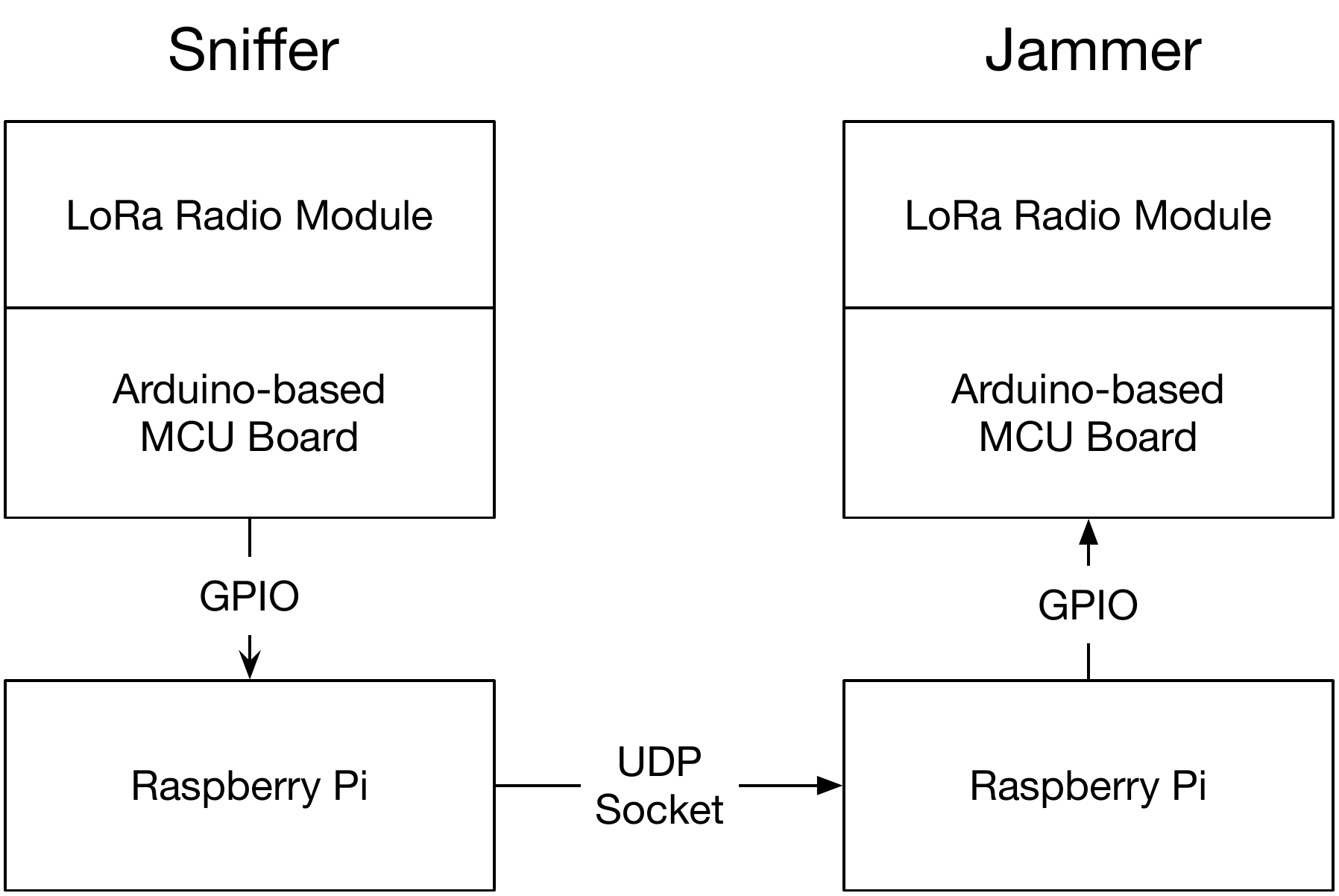}
  \caption{ Wormhole Attack Setup.}
  \label{fig:arch_worm}
\end{figure}

Figure \ref{fig:arch_worm} shows the architecture designed for this attack.
In this setup, we used the same radio modules and Arduino boards that we used for the selective jammer described in the previous section. There needs to be a low-latency link (i.e., faster than LoRaWAN) between the two devices to perform this attack correctly.
We decided to use UDP communications over Ethernet to connect these two devices together. There are some hardware options that can be found on the market that allow us to implement UDP communications such as Arduino Ethernet shields or embedded computers such as Raspberry Pi \footnote{https://www.raspberrypi.org}, or Beagle Bone \footnote{https://beagleboard.org}. We used two Raspberry Pi to create this UDP socket. One of them is connected to the sniffer and the other one is connected to the jammer via GPIO. Once the sniffer decides a LoRaWAN message needs to be jammed, it sets one of its external pins to a high state to assert an interrupt in the Raspberry Pi. The Raspberry Pi then sends UDP messages to the Raspberry Pi connected to the jammer. This second Raspberry Pi sets one of its external pins to a high state to assert an interrupt in the Arduino board. The Arduino board then turns the jammer on immediately.


\section{Evaluation}
\label{sec:evaluation}

\subsection{Traffic Analysis}
\label{sec:traffic}
As mentioned in Section \ref{sec:bg}, packet size affects the air-time of LoRa messages, which means that to test the effectiveness of the jammer it is important to calibrate the size of the test packets.
An analysis of LoRaWAN traffic was performed using a MultiTech Gateway \footnote{http://www.multitech.com/brands/multiconnect-conduit} and its logs to gauge the packet size of LoRa transmissions in regular use. We recorded LoRa traffic in two locations, one in Europe and one in the USA. The first, on the KU Leuven campus in Belgium, was performed over 6 days, totalling 1383 messages from 86 different device addresses bound for the local university gateway. These messages were spread across 8 channels, as follows (in MHz): 867.1 (9.1\%), 867.3 (10.7\%), 867.5 (8.9\%), 867.7 (10.3\%), 867.9 (9.6\%), 868.1 (17.1\%), 868.3 (18.7\%), and 868.5 (15.5\%).
The second, at an IoT event in Philadelphia recorded traffic from commercial devices for approximately two hours. The 87 recorded messages there were spread over 9 channels, as follows (in MHz): 911.9 (11.5\%), 912.1 (11.5\%), 912.3 (12.6\%), 912.5 (10.3\%), 912.6 (12.6\%), 912.7 (10.3\%), 912.9 (11.5\%), 913.1 (8\%), 913.3 (11.5\%).
The packets averaged 18.6 bytes in the first case, and 20.3 bytes in the second. Both of these are far lower than the maximum allowed packet sizes shown in Table \ref{tab:max_payload}. Considering 13 bytes of every message are consumed by headers, this means that the average payload visible from the gateway was 5-6 bytes in the first case, and 7-8 bytes in the second. To ensure that the jammer can jam the average sized packet, the slightly smaller test packet size of 17 bytes was used as a baseline in the following experiments.

\subsection{Selective Jamming}

In order to evaluate this attack, we used two MicroChip RN2483 Class A end-devices. To highlight the selective aspect, one of these device was the jamming target, while the other was used as a control, our aim being to only jam the first.
Table \ref{tab:selec} shows the messages received and jamming percentage at the gateway during the experiment. \emph{For each SF, 1000 messages were sent from each device, on the same frequency (868 MHz) enabling only one channel(868.1 MHz) on the end-devices. For all SFs, the jamming percentage is higher than 98\%.}

\begin{table}
  \caption{Selective jamming experiment}
  \label{tab:selec}
  \begin{tabularx}{\columnwidth}{P{0.2\columnwidth}|P{0.2\columnwidth}|P{0.2\columnwidth}|P{0.2\columnwidth}}
    & \multicolumn{2}{c|}{Packets received from:} & \\
    Spreading factor & Jammed device & Control device & Jam percentage \\
    \hline
    7 & 10 & 1000 & 99\% \\
    8 & 13 & 1000 & 98.7\% \\
    9 & 9 & 1000 & 99.1\% \\
    10 & 10 & 1000 & 99\% \\
    11 & 1 & 1000 & 99.9\% \\
    12 & 6 & 1000 & 99.4\% \\
  \end{tabularx}
\end{table}

\subsection{Combined Selective Jamming and Wormhole Attack}
As described in Section \ref{sec:attacks}, this attack is very time-sensitive, as the overhead of the wormhole communications have a big impact on the time to jam packets. To quantify that impact, we evaluated the attack in two parts. First, we measured the latency that is introduced by the hardware, the software stack, and the network between the sniffer and the jammer. The wormhole setup was timed end-to-end  by having the Arduinos at either end of the setup flash an LED after detection of a message to jam at the sniffer side, and at the jammer side when sending jamming messages. This setup was recorded fifteen times using high speed (240fps) video. The time between both flashes was then measured by counting the video frames between each flash. The mean observed time was 100.83ms, with a standard deviation of 1.7ms.


\begin{figure}
  \includegraphics[width=\columnwidth]{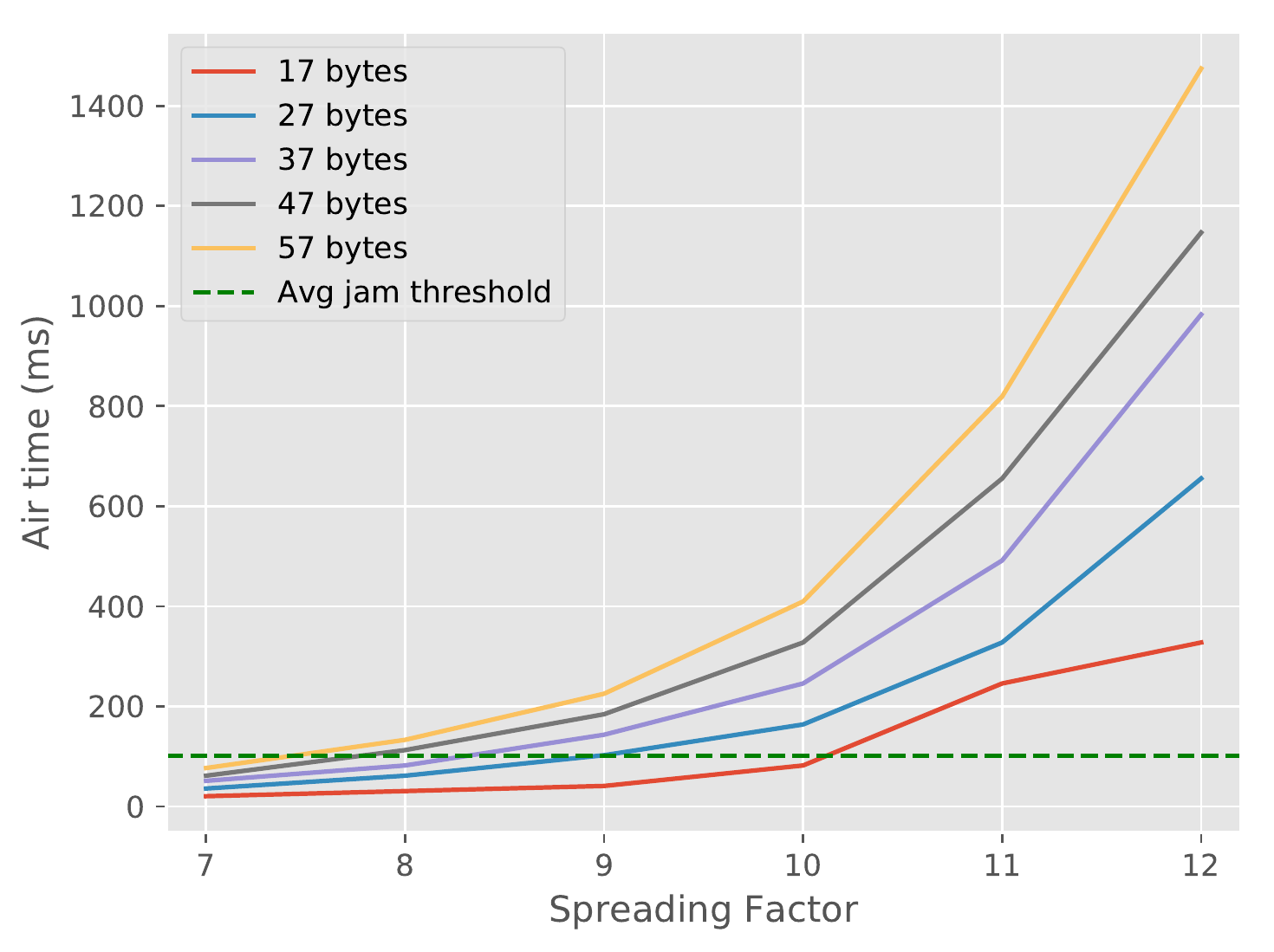}
  \caption{Time available to jam for various packet sizes for each SF. Times above the threshold should be reliably jammable.}
  \label{fig:e2eairtime}
\end{figure}

Using the formulas given in LoRa documentation \cite{semtech}, it is possible to calculate an approximation of the air time of any given LoRa message. By subtracting the given airtime of the first five bytes of a message (enough to read the device address) from the total airtime of a message, it is possible to calculate the jamming window of a message in the case of selective jamming. If that window exceeds the reaction time of the jammer, the message should be jammable. Figure \ref{fig:e2eairtime} shows this process applied to our jammer. Comparing the reaction time calculated above to the projected time-on-air of LoRa messages of increasing sizes (from 17, our baseline, up in increments of 10), allows us to predict how large a message would have to be to be jammable by our setup at lower SFs.

To quantitatively evaluate the accuracy of this model (i.e., what SF/packet size pairs are actually jammable), we used a MicroChip RN2483 module as jamming target, and sent 100 messages for each SF/packet size pairs close to the predicted threshold of the model described above. Table \ref{tab:worm_exp} shows the results of this experiment.


\begin{table}
  \caption{Wormhole attack experiment results. This table presents the capabilities of the Wormhole Jammer setup for each SF/Packet Size pair. \colorbox{green!30}{S}: successful jamming (>95\%), \colorbox{yellow!30}{M}: mixed success (0-95\%), \colorbox{red!30}{F}: failure to jam (<0\%).}
  \label{tab:worm_exp}
  \begin{tabularx}{0.7\columnwidth}{c|c|c|c|c|c|c|}
     & \multicolumn{6}{c|}{SF} \\ \cline{2-7}
    Message Size & 7 & 8 & 9 & 10 & 11 & 12 \\
   \cline{1-7}
  17 & \colorbox{red!30}{F} & \colorbox{red!30}{F} & \colorbox{red!30}{F} & \colorbox{yellow!30}{M} & \colorbox{green!30}{S} & \colorbox{green!30}{S} \\
  27 & \colorbox{red!30}{F} & \colorbox{red!30}{F} & \colorbox{yellow!30}{M} & \colorbox{green!30}{S} & \colorbox{green!30}{S} & \colorbox{green!30}{S} \\
  37 & \colorbox{red!30}{F} & \colorbox{red!30}{F} & \colorbox{yellow!30}{M} & \colorbox{green!30}{S} & \colorbox{green!30}{S} & \colorbox{green!30}{S} \\
  47 & \colorbox{red!30}{F} & \colorbox{red!30}{F} & \colorbox{green!30}{S} & \colorbox{green!30}{S} & \colorbox{green!30}{S} & \colorbox{green!30}{S} \\
  57 & \colorbox{red!30}{F} & \colorbox{red!30}{F} & \colorbox{green!30}{S} & \colorbox{green!30}{S} & \colorbox{green!30}{S} & \colorbox{green!30}{S} \\
  \end{tabularx}
\end{table}

As shown in Table \ref{tab:worm_exp}, the model correctly predicts that SF7 messages cannot be intercepted by this setup, and that SF11 and SF12 messages are all jammable. The real-world performance for SF10 approximates the predicted performance, with 27 byte packets being consistently jammable, although the 17 byte packets should be below the threshold but still get jammed in 40\% of cases. For SF9, the model predicts that 37-57 byte packets
should be jammable, 27 byte packets right on the threshold, and 17 byte packets not jammable. The actual results diverge somewhat, with 37 byte packets not consistently jammable (40\%), whereas the model predicts 40ms of jamming time. Finally, the model has SF8 packet sizes 47-57 above the jamming threshold , while the experiment showed that no SF8 packets were jammable.

It is worth noting that the LoRa time-on-air calculations are only approximations, and as shown with this experiment can only be used to gain a rough estimate of the likelihood of jamming a packet when the prediction is close to the threshold. This model is appropriate when considering bigger gaps however (> 40-50ms in our case).


\section{Applicability to Real-World Scenarios}
\label{sec:real}
\subsection{Attack Limitations}
\label{sec:limits}

The jamming attacks presented in this paper rely on two characteristics of LoRa messages: i) the long air-time of these messages, which affords the time to react to their presence, ii) and the possibility of drowning out legitimate messages with jamming messages broadcast with more power. To be successful, a jamming attack will need to coordinate these two, which may be difficult in some cases.

\subsubsection{\bf Window of Opportunity}

All of the jamming attacks presented in this paper rely on a detection and reaction mechanism to jam incoming messages. The selective jamming attack increases the length of the detection time (compared to triggered jamming) by having to read more of the incoming messages before triggering the jamming. The wormhole attack adds a significant additional delay to the reaction part of the process. As shown in Section \ref{sec:wormhole}, this delay prevented our wormhole attack from jamming lower SFs, especially at low packet sizes.
The setup used in the experiment relied on a cabled Ethernet as the low-latency link, a best case scenario. In a real-case, the wormhole attack will likely need to use a long-range fast wireless link, such as WiFi internet. To establish the impact of the use of WiFi, the model presented in Section \ref{sec:wormhole} can be used. The accuracy of the model was tested against the real-world efficacy of the setup, allowing the model to be used in a more general manner, to predict how usable alternative communication technologies would be. By adding the networking time expected for these alternative technologies to the setup's built-in latency, we can predict whether or not the technology is usable for a particular SF and message size. The overhead of WiFi internet (~10ms) would still enable jamming of most higher SF cases.


Its is important to note that the experiments conducted in this paper were done with commodity hardware with a total cost of approximately 100 euros. The latency of the setup could be reduced by writing a custom RaspberryPi driver or using more expensive and dedicated hardware (e.g., FPGA boards), improving the success rate at lower SFs.

\subsubsection{\bf Jammer Signal Strength}

\begin{figure}
  \includegraphics[width=\columnwidth]{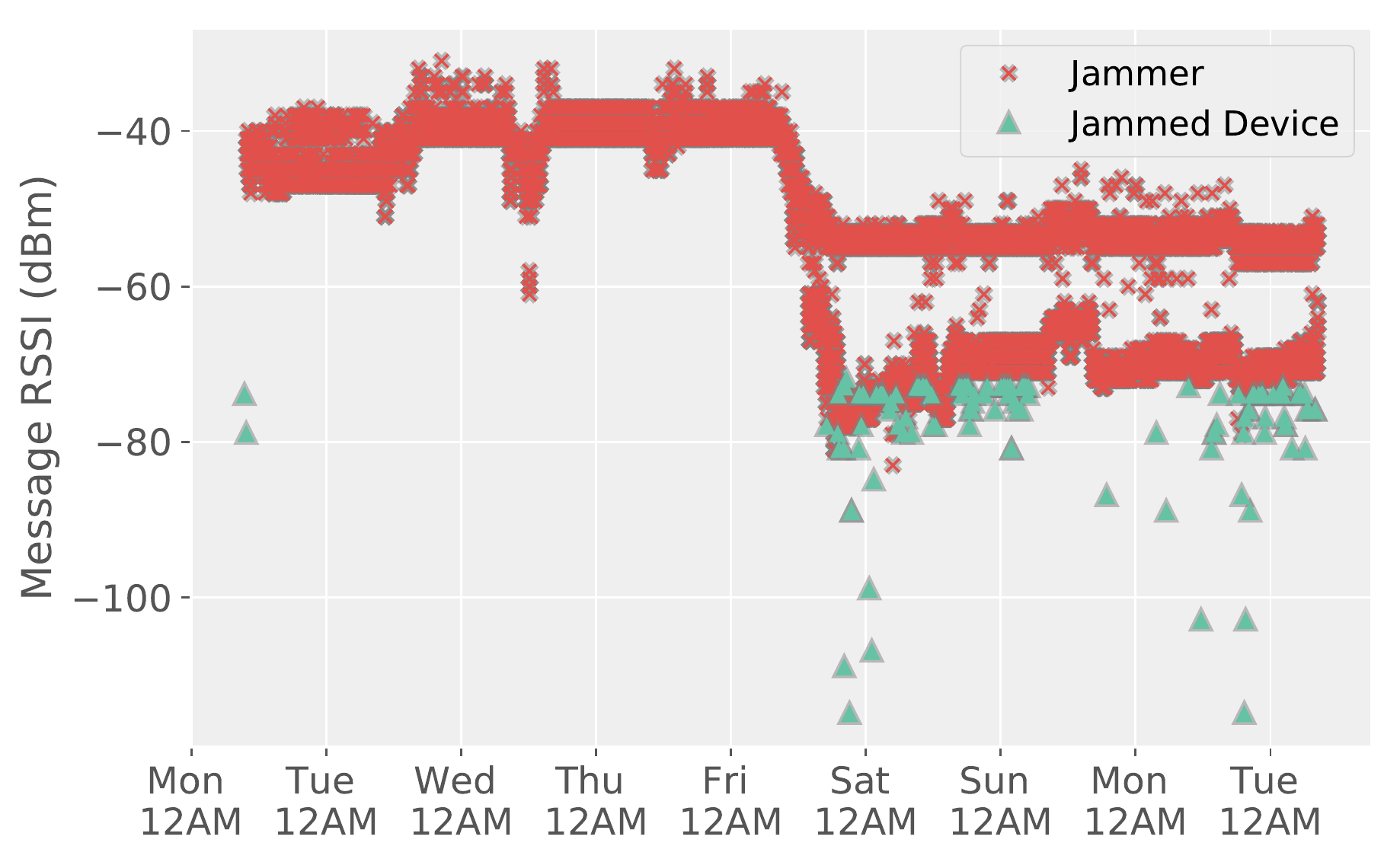}
  \caption{Effects of RSSI drop on jamming attempts.}
  \label{fig:rssi_drop}
\end{figure}

Signal strength is an important factor in the ability of the jammer to jam a device. If, at the gateway, the RSSI (Received Signal Strength Indication) of the jammer is too low relative to the jammed device, the jamming attempt will fail regardless of timing issues (See Figure \ref{fig:db_thresh} for the relevant thresholds). Figure \ref{fig:rssi_drop} shows the results of a jamming experiment conducted at SF12. The figure shows the recorded RSSI of the Jammer and the Jammed Device over time. According to \cite{goursaud_dedicated_2015}, the calculated threshold differential between jammer and source for jamming at SF12 is 36dB, a value which is aligned with the results shown in Figure \ref{fig:rssi_drop}. When the Jammer's RSSI is around -40, almost all packets are jammed, whereas when this RSSI dropped to around -60, some packets leak through, with the jammed device's RSSI recorded at around -80 throughout. The RSSI of the jamming messages at the gateway can be increased in three ways: 1) by moving the jammer closer to the gateway, 2) increasing the output power of the jammer, and 3) through the use of directional antennas.

\subsubsection{\bf Channel Hopping}

The attacks presented in the previous sections make use of single channel hardware: the channel to listen and broadcast on is set before doing either. This is in contrast to gateway hardware for example, which is able to listen to multiple channels simultaneously. For our tests, the jammed device was constrained to sending on a single channel, on which the jammer was set to listen/broadcast. As shown in Section \ref{sec:traffic} however, regular LoRa devices hop between channels each time they send a message to avoid accidental collisions. To implement this attack in a real scenario would require listening for messages on multiple channels to know where to jam. This is possible through the use of a multi-channel receiver such as found in LoRa gateways, or by using multiple single channel chips such as those use in the current setup. However this will moderately increase the price of the required hardware.

\subsection{Potential Targets}




When discussing potential targets and real-world applications of these attacks, it is important to note that the selective jamming and the wormhole attack described in the previous sections are designed to jam the gateway in order to prevent communications coming from an end-device to reach a gateway.
If the scenario is reversed (i.e., jamming gateway to device transmissions), the process would change to waiting for an end-device to start a transmission sequence, then relying on a timer, jam the two subsequent receive slots as their timing sequence is fixed in the specifications and therefore predictable.
This section explores the potential of each of the attacks described in this paper, suggesting the best use for each technique.

\subsubsection{\bf Triggered Jamming.}

In the triggered jamming case, the jamming is applied as a blunt obfuscation technique, meaning that any time the preamble of a LoRa transmission is detected, the jammer starts to saturate the input stage of the receiver in order to prevent any possible reception for the targeted gateway.
This attack is well suited to a massive perturbation of a set of gateways, disrupting the running applications and the network indiscriminately.
While it jams any LoRa transmission, this technique is far from being covert from the point of view of the network and application infrastructure, as any application-level monitoring should be able to detect this disturbance.

\subsubsection{\bf Selective Jamming.}

The selective jamming attack is instead directed towards a particular set of end-devices, the goal being to activate the jamming of messages coming from a particular end-device at a particular time.
The selectiveness of the attack makes it suitable for preventing a specific event (e.g., changes in sensor data) from being communicated to the gateway. This is especially effective against any device that sends messages only when a trigger occurs, such as alarms might. Such an attack may spoil the blood and vaccine supplies in case of medical fridge LoRa deployment in Kikwit, DR Congo~\cite{7945359}.
While the above describes jamming of a specific device, this selective jamming can be done on any of the headers in a LoRaWAN packet, as these are all broadcast in clear. For example, jamming based on message type can have severe impacts on a network, as for example all ``join'' messages could be jammed, preventing all devices from joining or re-joining the network.

\subsubsection{\bf Combined Selective Jamming \& Wormhole Attack.}

The combined selective and wormhole jammer attack adds another dimension to selective jamming by adding message replay capability. The replay requires additional setup (i.e., recording normal operational messages ahead of time), but enables more sophisticated attacks than selective jamming can achieve. By replaying normal operation messages while jamming, the attacker can not only intercept alert messages, but also make it seem as though nothing out of the ordinary is happening.

These attacks can be combined for maximum effectiveness; while the selective and wormhole attack is powerful, as shown in Section \ref{sec:evaluation} it can only act reliably on high SF communications. By using a triggered or selective jammer at a gateway close to a device to block join attempts with that gateway, it is possible to force a device to join a more distant gateway, forcing it to move to higher SFs, and therefore enabling the selective and wormhole attack to operate.

\section{Mitigation Techniques}
\label{sec:mitigation}
As discussed in Section \ref{sec:limits}, the attacks presented in this paper have limitations which can be used against them. This section describes a series of mitigation techniques that exploit these limitations and the characteristic changes caused by these attacks, to recognise and stop them.

\subsection{Low-Level Techniques}

By using the capabilities of LoRa itself, it is possible to reduce the effectiveness of the jamming techniques described above. The following list highlights possible approaches.

\begin{itemize}
  \item{\bf Create dense LoRa networks with overlapping coverage regions:} By deploying LoRaWAN end-devices within the range of multiple gateways, we increase the reliability of LoRa communication. This feature is critical in beating jamming attacks, as to guarantee that a message is jammed, the jammer must ensure it is heard at no gateway in the network. Since the jammer requires high RSSI compared to the end-device, the jammer is more effective when it is close to the gateway. Thus, the jamming is more complex in the presence of multiple gateways, as the malicious attackers must map the gateways in range of each target end-device to successfully jam the transmissions.


  \item{\bf Maximise the use of channel hopping.} As described in Section \ref{sec:limits}, LoRa devices hop between multiple channels when sending messages as dictated by LoRaWAN specification, to reduce the chance of collisions. The more channels used, the more complex the jammer has to be, as it needs to listen on all of those channels, forcing a move from basic low-cost LoRa hardware, to more expensive multi-channel LoRa receivers as found in gateways.

  \item{\bf Move to higher spreading factor (i.e., SF12) to beat jammer RSSI.} As pictured in Figure \ref{fig:db_thresh}, the higher SFs require higher dB differentials between the jammer and target message. Although higher spreading factor transmissions afford more time for the jammer to act, it still requires the jammer to be closer to the gateway. Note that multiple transmissions in higher SF quickly exhaust the duty cycle allowance.

  \item{\bf Beat jammer reaction time (Wormhole Only):}
  \begin{itemize}
    \item {\bf Move to low SF to beat jammer reaction time.} Reducing SF reduces the airtime of messages, which in turn reduces the time the jammer has to react. This has several costs however: i) Lower SFs have lower reliability and lower range, and ii) Lower SFs require less power output from the jammer to be disrupted.
    \item {\bf Drop packet size to beat jammer reaction time.} As shown in Figures \ref{fig:airtime} and \ref{fig:e2eairtime}, packet size has a significant impact on message air time. Reducing the size of these messages could allow messages to beat the jammer's reaction time.
  \end{itemize}
\end{itemize}

\subsection{Application-Level Techniques}
By performing traffic analysis and profiling (at the gateway or server level), it is possible to identify variations in the pattern of incoming messages indicating the presence of a jammer, and to trigger alarms or adaptations to the network. The following list highlights potential approaches based on differing assumptions:

\begin{itemize}
  \item {\bf When the transmission rate is known.} If the entity doing the traffic analysis is aware of the sending rate of the LoRa end-devices, it can easily identify unplanned changes in traffic patterns, and react accordingly.
  \item {\bf When the transmission rate is unknown.} In this case, the \emph{normal} rate of traffic needs to be established over time, or through past continuous profiling. Once the baseline rate is understood, it becomes possible to identify deviations.
\end{itemize}
In essence, the continuous monitoring and profiling of LoRa end-devices is essential to prevent security attacks on LoRa end-devices. LoRa gateways must incorporate mechanisms to allocate suitable spreading factor for end-devices along with time synchronisation and channel hopping to improve the security and the robustness of large scale LoRa deployments.

\section{Conclusion}
\label{sec:conclusion}
This paper showed that LoRa and LoRaWAN are susceptible to jamming attacks using low-cost, commodity hardware broadcasting normal LoRa messages.
Our work introduces a series of jamming attacks on LoRa traffic, that exploit the weakness of LoRa messages to interference caused by other synchronous LoRa messages of higher signal strength.
These attacks are enabled thanks to the long air-time of LoRa messages, which allows for the detection of a particular message and the emission of a jamming message while the original message is still being broadcast.
This paper presented a selective jamming attack which reads enough of a packet to assert its origin before jamming, and an attack combining selective jamming and a wormhole attack that splits the jammer over two devices connected by an alternative network.
This setup allows an end-device's messages to be read and recorded close to the source, while the remote node jams the message close to the gateway.
The recorded messages can then be played back in a delayed fashion to make the device appear to be functioning normally.
These attacks were tested in a real LoRa testbed and proven to function.
We discussed the applicability and inherent real-world limitations of these attacks, and proposed mitigation techniques, taking place both at the application level and low level of a LoRaWAN infrastructure.

Future work will investigate the mitigation strategies described in Section \ref{sec:mitigation}, and assess the effects of these attacks on the newer classes of LoRa devices (B and C) when they are released. Finally, we will
explore the possibilities of cryptographic channel hopping for LoRa, in order to render jamming attacks ineffective.


%

\section*{Acknowledgments}
This research is partially funded by the research fund KU Leuven and is conducted in the context of the IMEC HI\textsuperscript{2}-NETSEC Project. Support has been provided by VersaSense, provider of industrial IoT solutions (www.versasense.com).

\bibliographystyle{ACM-Reference-Format}
\bibliography{lora_wormhole_paper}

\end{document}